\documentclass[english,12pt]{article}
\usepackage[T1]{fontenc}
\usepackage[latin9]{inputenc}
\usepackage{graphicx}

\makeatletter
\newcommand{\lyxaddress}[1]{
\par {\raggedright #1
\vspace{1.4em}
\noindent\par}
}

\usepackage[style=science,natbib=true,backend=bibtex8,sorting=none]{biblatex}
\usepackage[a4paper, margin=1in]{geometry}
\usepackage{setspace}
\AtEveryBibitem{%
  \clearfield{issue}%
  }

\makeatother

\usepackage{babel}
\begin{document}

\title{Spatio-temporal dynamics of outbreak on a lattice with quenched mobility
patterns }

\author{Jozef \v{C}ern\'{a}k\footnote{jozefcernak@gmail.com}}
\maketitle

\lyxaddress{Department of Nuclear and Sub-Nuclear Physics, Faculty of Science,
Institute of Physics, Pavol Jozef Šafárik University in Košice, Košice,
Slovak Republic}
\begin{abstract}
We have designed a computational model of a virus spread near the
outbreak threshold. Using computer simulation we studied the Susceptible
- Infected - Recovered (SIR) process where in consequence of a force
of habit that is manifested by the population mobility patterns, the
recovered persons create the spatio-temporal patterns as the barriers
to a virus transmission. The results show a spontaneous stopping of
the virus spread without a need to infect the whole population, a
non-trivial random noise of daily count of infected cases, and power
laws of a cumulative count of infected cases. Outbreak evolution strongly
depends on the initial conditions thus we concluded that the model
has the features of chaotic systems that makes it difficult to predict
its behaviors. 
\end{abstract}

\section*{Introduction}

The SARS-CoV-2 virus and their variants continue in the evolution
in the global space. The destiny of virus extinction is unclear \cite{Shaman527}
and it depends on many evolutionary factors \cite{Callaway}. An increase
of social distance is a measure to slow down the pandemic. It has
been successfully demonstrated that either local or national non-pharmaceutical
interventions led to a significant reduction of the virus rate transmission
on large scales with beneficial and measurable health outcomes \cite{Hsiang2020, Kissler860}. 

The mathematical models, susceptible-infected-recovered (SIR), susceptible
- infected -susceptible (SIS), susceptible - infected - recovered
- susceptible (SIRS) and susceptible - exposed - infected - recovered
-susceptible (SEIRS) \cite{Pastor_Satorras, Newman_2018}, predict
an exponential growth of a cumulative count of infected cases $I(t)\sim\exp(t)$,
where $t$ is time. Similarly, the SI, SIS, and SIR models on the
complex networks show an exponential growth of the number of links
available for future transmission \cite{Barabasi_2016}. These theoretical
predictions are in contrast to a cumulative count of infected cases
$I(t)$ in China during the beginning of the first SARS-CoV-2 outbreak
wave. The first wave in China shows a power law, $I(t)\sim t^{\alpha}$
\cite{Singer_2020}, where $\alpha$ is an exponent. A deviation from
the expected exponential growth models, we consider for a signature
of a quite different mechanism of the contagion, than it is widely
accepted for traditional SIR or SEIR models \cite{Lai2020}.

The authors Rhodes and Anderson \cite{Rhodes1996} analyzed distribution
of epidemic sizes and epidemic durations of measles outbreaks. They
observed that dynamical structures of the measles returns reflect
the existence of an underlying scaling mechanism. Random structures
often exhibit self-similar geometry that is characterized by a power
law \cite{ORBACH814}. So, it is useful to consider a virus contagion
as a dynamical process on the fractal networks like a diffusion in
the percolating networks \cite{ORBACH814} as well as the branching
process \cite{Antia2003} on these networks. Kumamoto and Kamihigashi
\cite{Kumamoto} reviewed mathematical mechanisms that are known to
generate the power laws. In particular, they focused on stochastic
processes based on growth and preferential attachments including the
Yule process, the Simon process, the Barab\'{a}si - Albert Model,
and stochastic models based on geometric Brownian motion.

Stroud et al. \cite{STROUD2006301} introduced the power laws of some
variables in the traditional homogeneous models, SIR and SEIR, to
better model the real outbreaks. In the stochastic version of the
SIR process, the authors Ben-Naim and Krapivsky \cite{Ben_Naim} found
nontrivial scaling relations of a maximal size of outbreak $M\sim N^{\frac{2}{3}}$
and duration of outbreak $T\sim T^{\frac{1}{2}}$ on a population
size $N$ near the outbreak threshold.

We observed that the time series of daily count of infected cases
$i(t)$ show a noise that is not possible to neither reproduce nor
explain by the classical models of epidemic. Hurst investigated annual
values of some phenomena such as river discharges, rainfall, temperatures.
He observed that these values are approximately normally distributed
if no account is taken of order of occurrence. So far as is known,
there is no regularity in the occurrence or the length of these periods,
and usually there is no significant correlation over one of them between
a year and its successor. Hurst considered this phenomenon to be important
in problems of storage \cite{HURST1957}. Mandelbrot found that the
Hurst exponent $H$ must be $0\leq H\leq1$. If $H<\frac{1}{2}$ then
a noise shows an anti-persistent fractional Brownian motion \cite{Mandelbrot_1983}.
If $H>\frac{1}{2}$ then the noise exhibits a long term persistence
and nonperiodic cycles. The Hurst noise was observed in many natural
phenomena \cite{OConnell2016} and it was demonstrated in the laboratory
insect populations \cite{Miramontes_1998}, i.e. the Hurst phenomenon
was reproduced in well controlled experimental conditions.

Our motivation is to better understand the emergence of power laws
in the outbreak evolution, the mechanisms of spontaneous stopping
of the outbreak and the nature of the noise of the daily count of
infected cases $i(t)$. We were inspired by the cellular automata
\cite{Wolfram1984}, a few features of the self organized criticality
models and forest fire models \cite{Bak,Malamud, Pruessner_2012}.
However, we had to constrain mobility of individuals and to introduce
a spatio-temporal memory effect.

\section*{Experimental results\label{sec:Experimental_Results} }

\begin{figure}
\centering

\includegraphics[width=8cm]{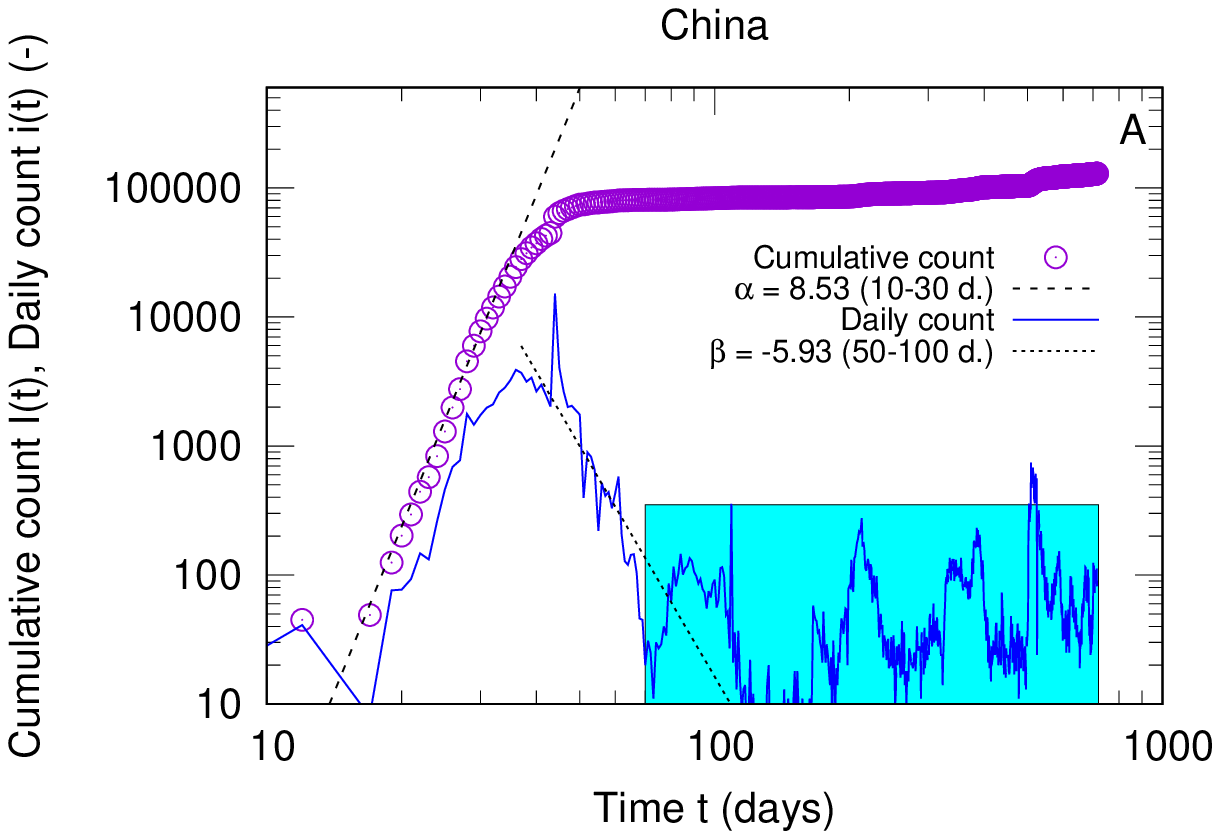}\includegraphics[width=8cm]{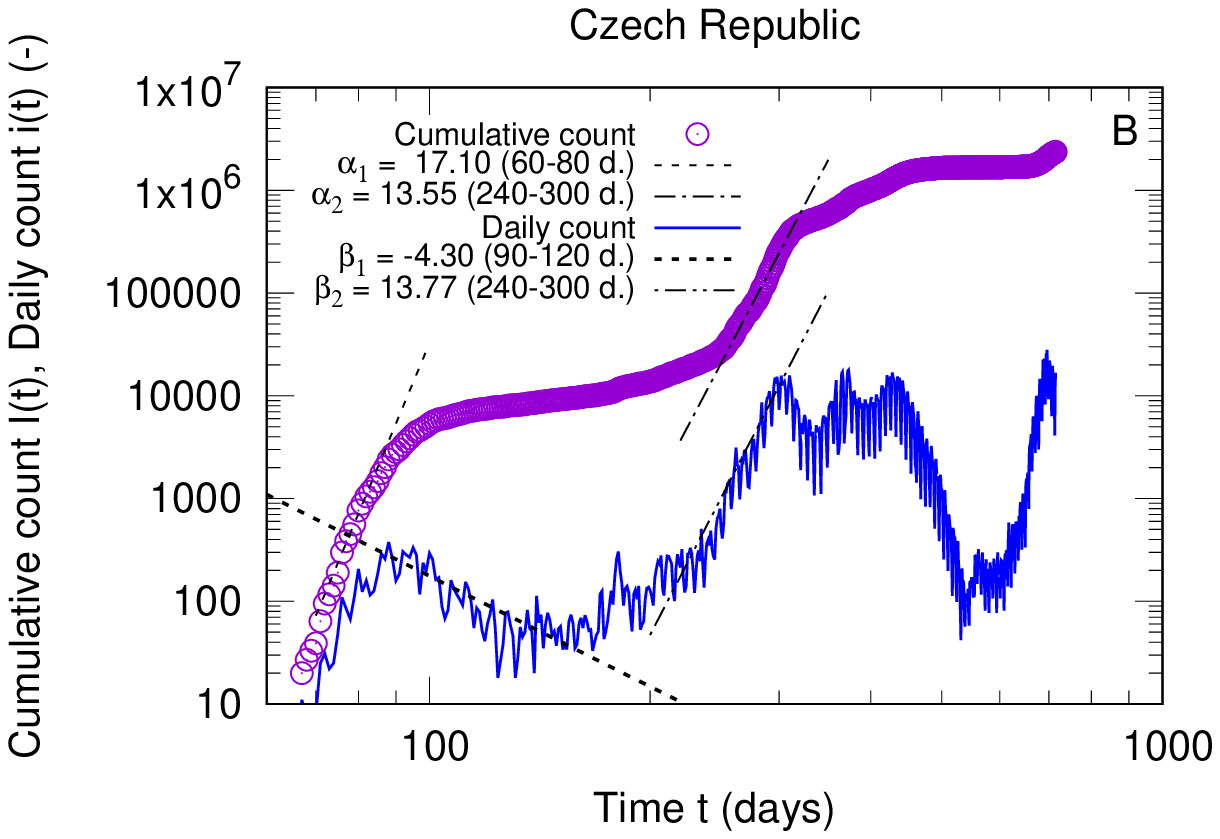}

\includegraphics[width=8cm]{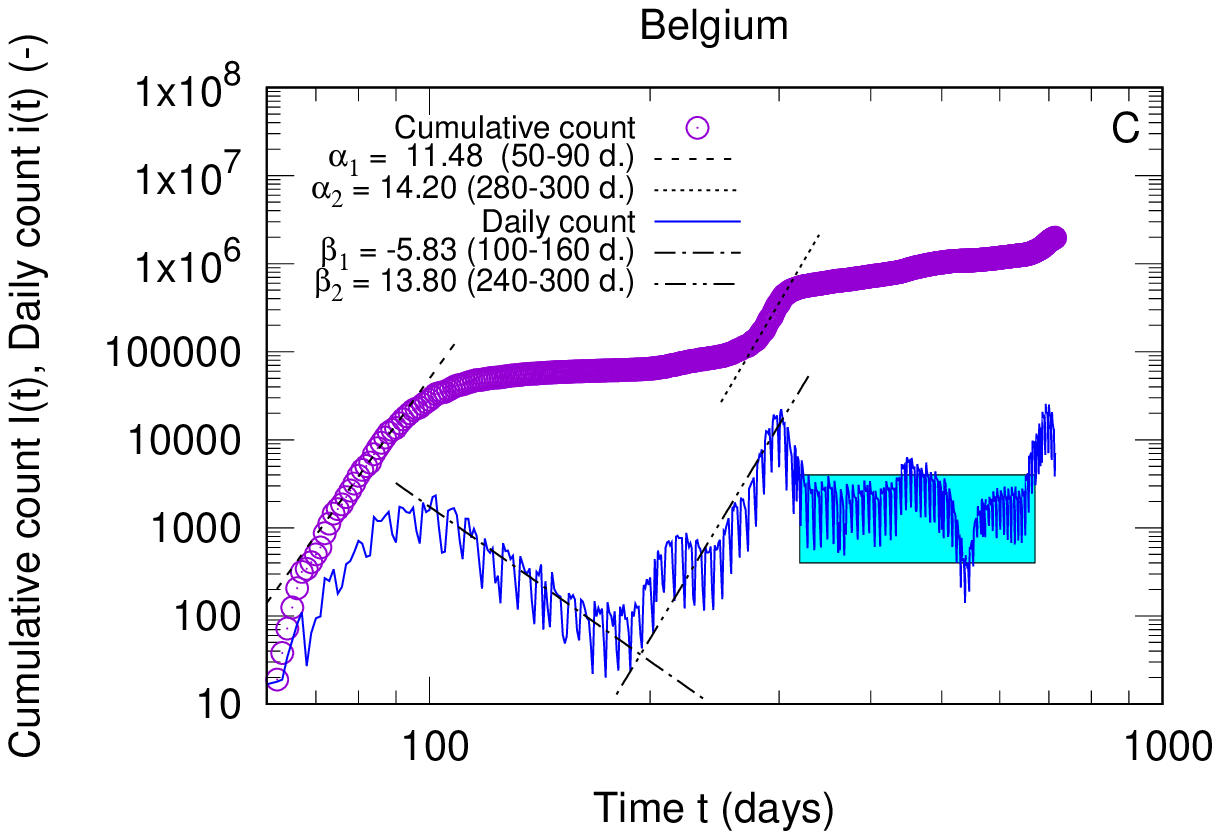}\includegraphics[width=8cm]{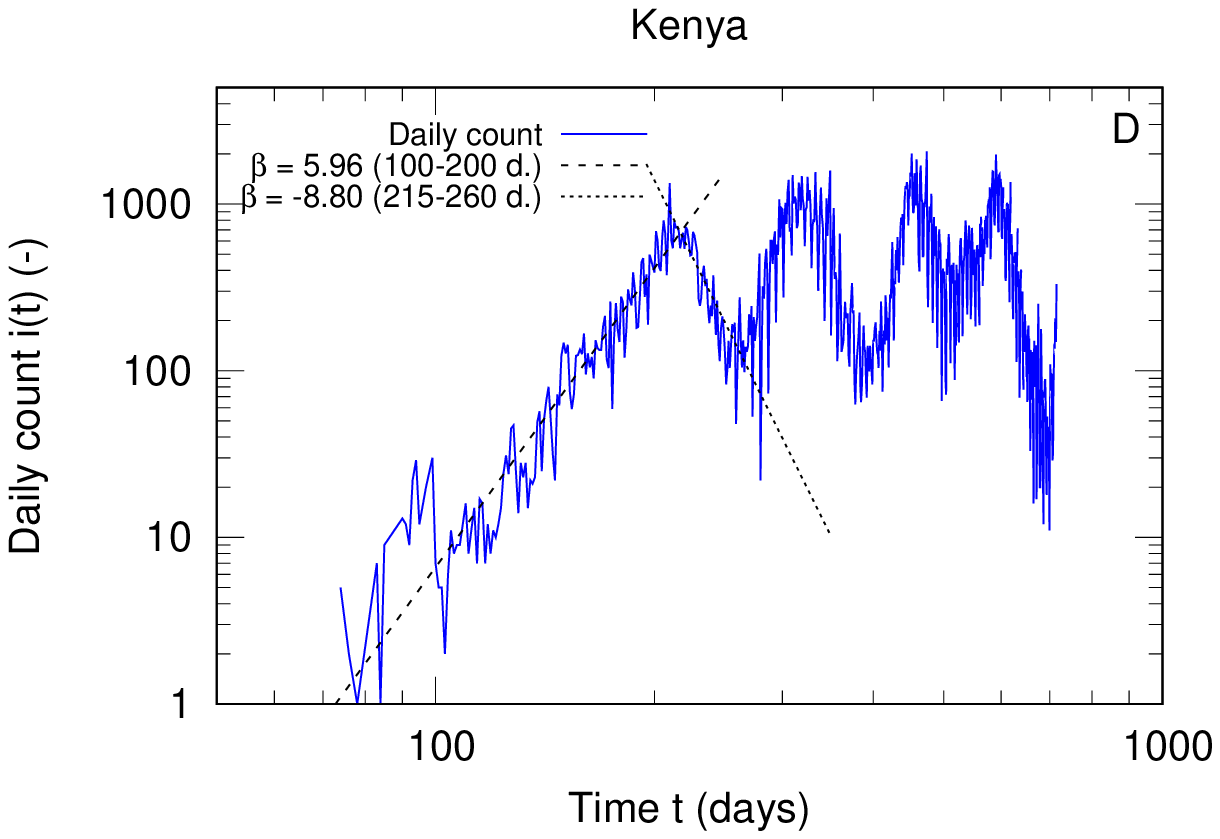}

\caption{\label{fig:experimental}Power law growth of the cumulative count
of infected cases $I(t)\sim t^{\alpha}$ and daily count of infected
cases $i(t)\sim t^{\beta}$, where the exponents are: (A) $\alpha=8.53$$\pm0.40$,
and $\beta=-5.93\pm0.80$, (B) $\alpha_{1}=17.10\pm0.79$, $\alpha_{2}=13.53\pm0.25$
$\beta_{1}=-4.30\pm0.82$ and $\beta_{2}=13.77\pm1.14$, (C) $\alpha_{1}=11.48$$\pm0.28$
$\alpha_{2}=14.20\pm0.25$ $\beta_{1}=-5.83\pm0.63$ and $\beta_{2}=13.80\pm1.60$
and (D) $\beta_{1}=5.96\pm0.38$ and $\beta_{2}=-8.80\pm0.8$. }
\end{figure}

World Health Organization (WHO) provides the SARS-CoV-2 data (Supplementary
Materials). We carefully selected countries: China, the Czech Republic,
Belgium and Kenya (Figure \ref{fig:experimental}), that show general
as well as country specific features of the virus spread. 

The daily count of infected cases $i(t)$ and the cumulative number
of infected cases $I(t)$ (Figure \ref{fig:experimental}) follow
the power laws $i(t)\sim t^{\beta}$and $I(t)\sim t^{\alpha},$ where
$t$ is a time measured in days and $\alpha$ and $\beta$ are exponents.
We observed a power law growth that should be directly linked to either
a decrease of mobility or other measures to prevent the virus spread.
If an outbreak grows, the exponents are $5.96\leq\beta\leq13.80$.
If the outbreak intenzity declines (an outbreak decay) then the exponents
are$-8.80\leq\beta\leq-5.93$.

Outbreak dynamics in China (Figure \ref{fig:experimental} (A)) is
country specific. Daily counts of infected cases $i(t)$ are low and
persist to fluctuate, for a long time. This type of dynamics prevents
the uncontrolled growth of outbreak size $M$. Rare character of the
outbreak dynamics in China, i.e. these long term fluctuations of small
daily counts of infected cases $i(t),$ is impossible to explain using
the traditional contagion models \cite{Pastor_Satorras, Newman_2018, Barabasi_2016}.

Rare time series of daily count of infected cases $i(t)$ (Figure
\ref{fig:experimental}) are limited by the time windows of $i(t).$
Similar rare fluctuations of $i(t)$, but fluctuating around a higher
value than in the China, were observed in Belgium (Figure \ref{fig:experimental}
(C)) for a long time period, more than $100$ days. Generally, fluctuations
of daily count of infected cases $(t)$ are typical for all countries,
for example in the Czech Republic (Figure \ref{fig:experimental}
(B)) and in Kenya (Figure \ref{fig:experimental} (D)) where the fluctuations
of $i(t)$ are superimposed on outbreak waves. 

We evaluated the Hurst exponent $H$ of the fluctuations of $i(t)$
for both experimental (Figure \ref{fig:experimental}) and computational
(Figures \ref{fig:simulation_1} and \ref{fig:simulation_2}) time
series \cite{Miramontes_1998, OConnell2016} (Supplementary Materials).
The Hurst exponents $H$, for experimental time series of daily count
of infected cases $i(t)$ (Figure \ref{fig:experimental}), were found
as: Figure \ref{fig:experimental} (A) $0.56\leq H\leq0.59$, Figure
\ref{fig:experimental} (B) $0.28\leq H\leq0.35$, Figure \ref{fig:experimental}
(C) $0.25\leq H\leq0.44$ and Figure \ref{fig:experimental} (D) $0.29\leq H\leq0.33$.
In the time windows where the daily count of infected cases $i(t)$
fluctuates around a certain value, we determined the Hurst exponents
$H$: Figure \ref{fig:experimental} (B) $H=0.28$, for time $300\leq t\leq500$,
Figure \ref{fig:experimental} (C) $H=0.25$, for time $350\leq t$$\leq670$,
and Figure \ref{fig:experimental} (D) $H=0.32$, for time $200\leq t\leq670$. 

\section*{Results of computer simulations}

\begin{figure}
\centering 

\includegraphics[width=8cm]{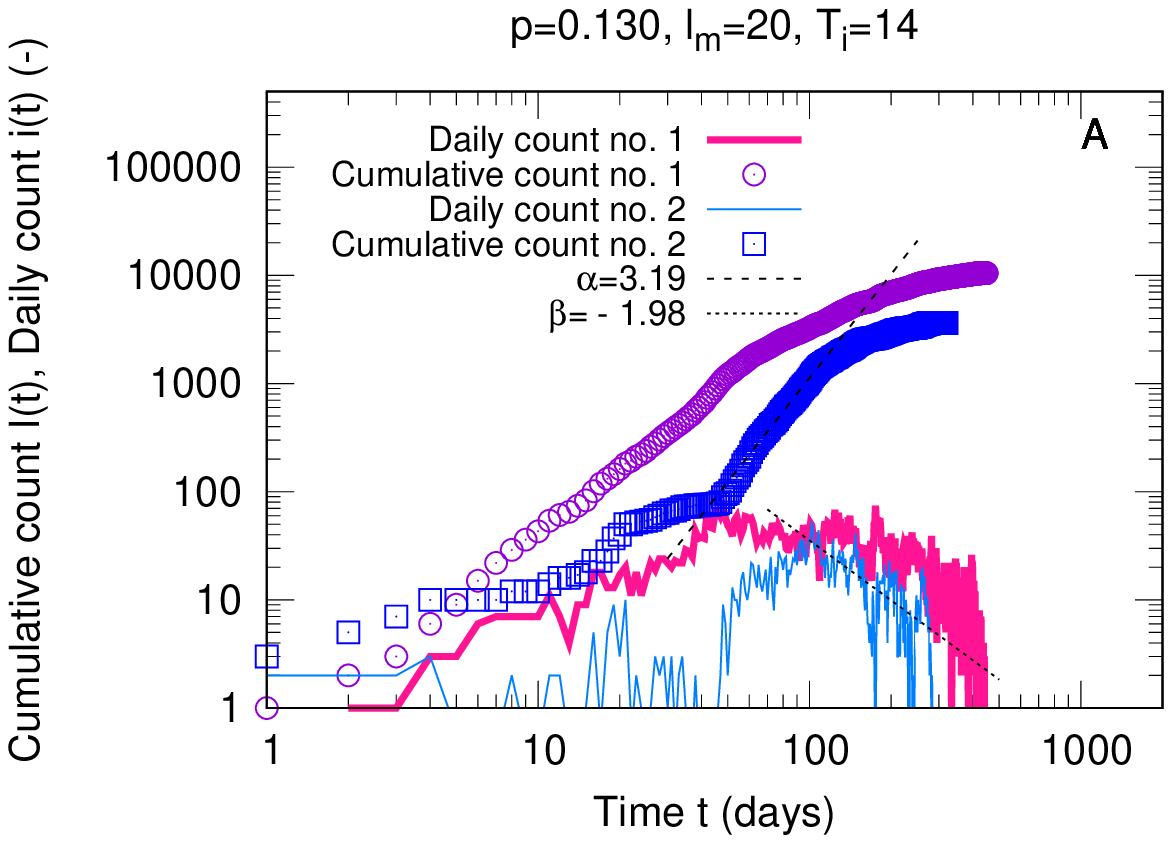}\includegraphics[width=8cm]{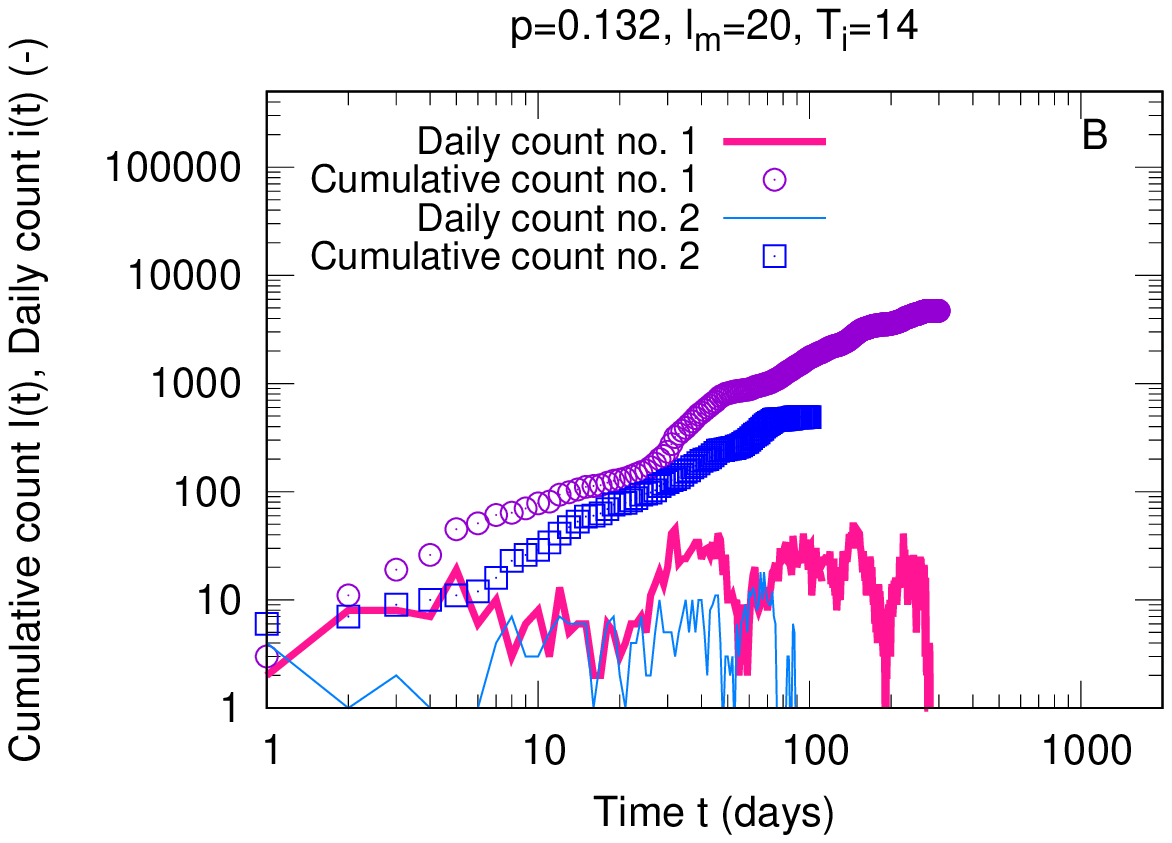}

\includegraphics[width=8cm]{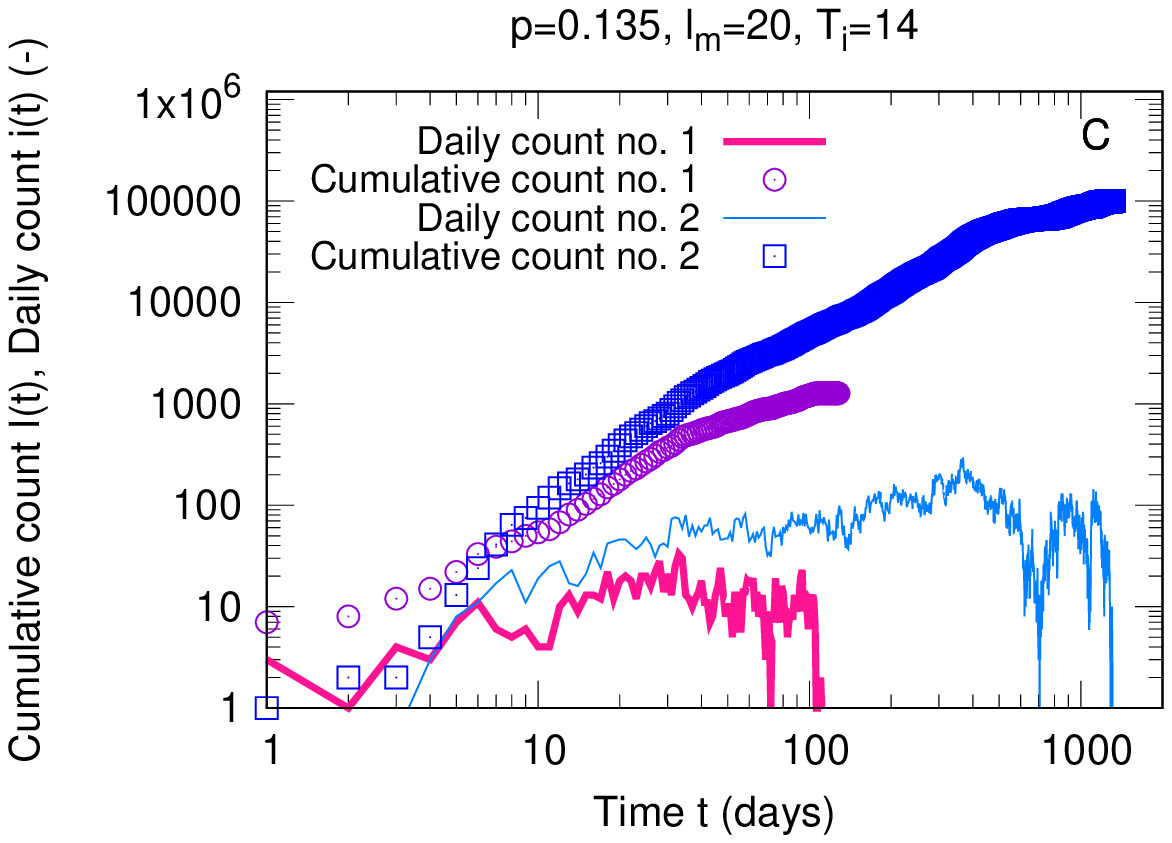}\includegraphics[width=8cm]{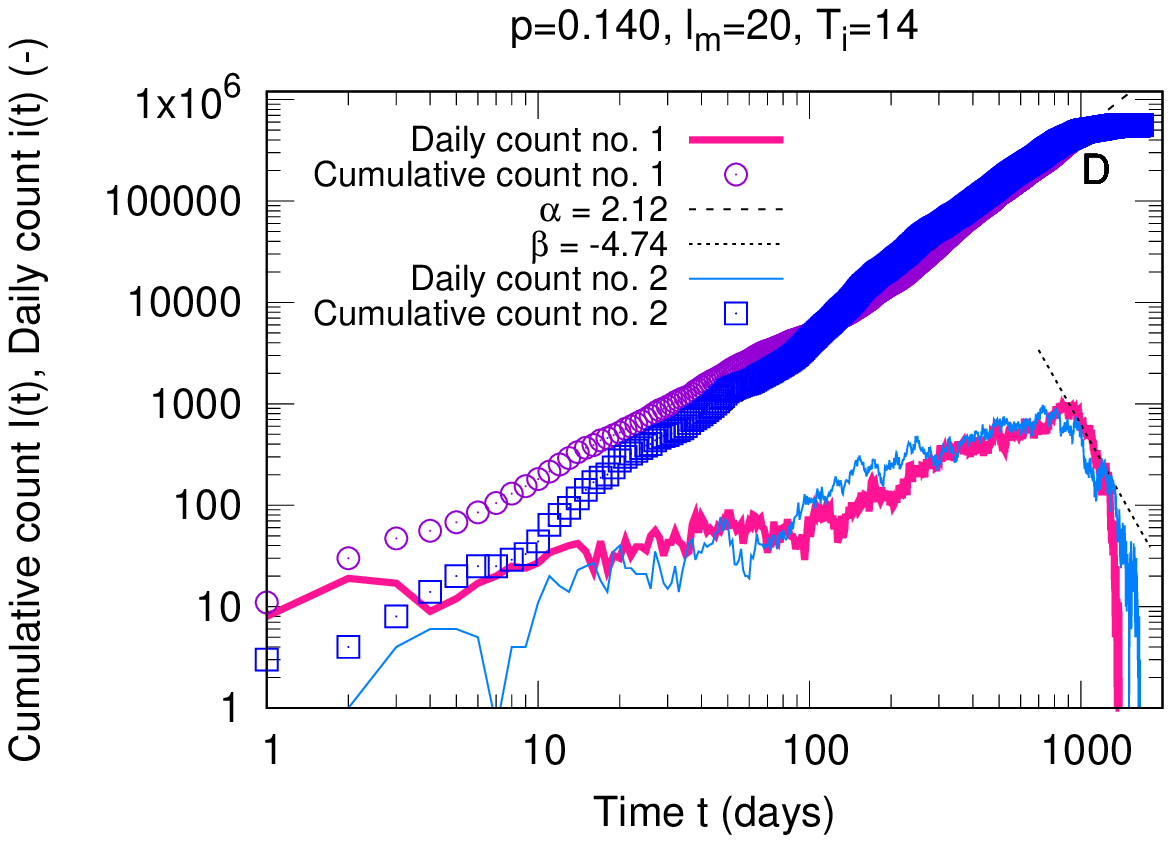}

\caption{\label{fig:simulation_1} The computer simulations of the cumulative
count of infected cases $I(t)\sim t^{\alpha}$ and daily count of
infected cases $i(t)\sim t^{\beta}$, for the variable parameter $p$,
$0.130\leq p\leq0.140$, and constant parameters $T_{i}=14$ and $l_{m}=20$.}
\end{figure}

The computer simulations, for the constant parameters $T_{i}=14$
and $l_{m}=20$, and the variable parameter $p$, $0.130\leq p\leq0.140$,
(Figure \ref{fig:simulation_1}) show a power law of the cumulative
count of infected cases $I(t)$, $I(t)\sim t^{\alpha}$ with the exponents
$\alpha=2.12$ and $\alpha=3.19$ (Figures \ref{fig:simulation_1}
(A) and (D)). We observed the outbreak decline (an outbreak decay)
that follows the power law of the daily count of infected cases $i(t)\sim t^{\beta},$
with exponents $\beta=-1.98$ and $\beta=-4.74$ ((Figures \ref{fig:simulation_1}
(A) and (D))).

Important finding is that the outbreaks can spontaneously stop their
growth after a time $T$. We are pointing up that spontaneous stopping
of outbreaks occur without a need to infect the whole population,
i.e. $M\ll N.$ In this growth regime, if $p<0.140$ ((Figure \ref{fig:simulation_1}
(A)-(C))), the duration of outbreak $T$ and maximal count of infected
cases $M$ are random variables that span a wide interval. An outbreak
(Figure \ref{fig:simulation_1} (A)-(C)) may stop to grow ($i(t)=0$)
and restart to grow again ($i(t)>0$). On the other hand, the restarts
of the outbreak are rare if the probability of virus transmission
is $p=0.140$ Figure \ref{fig:simulation_1} (D).

\begin{figure}
\centering

\includegraphics[width=8cm]{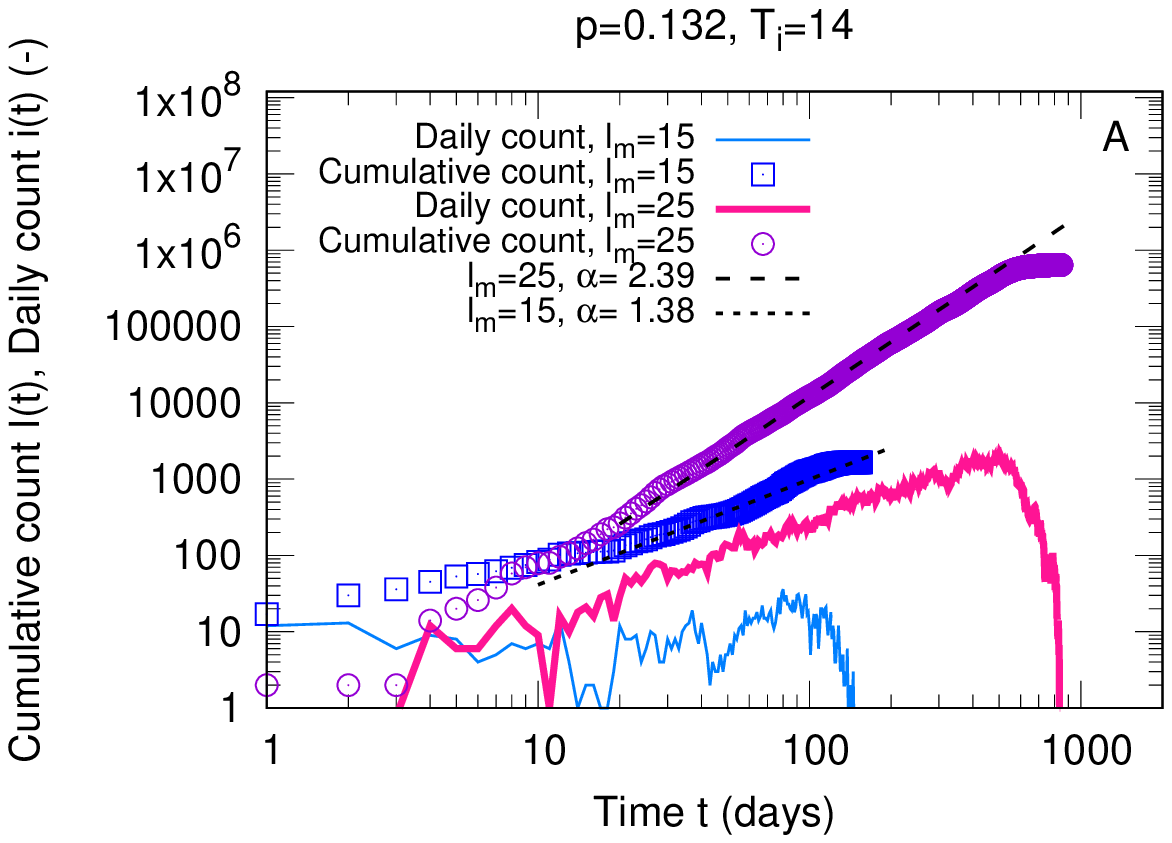}\includegraphics[width=8cm]{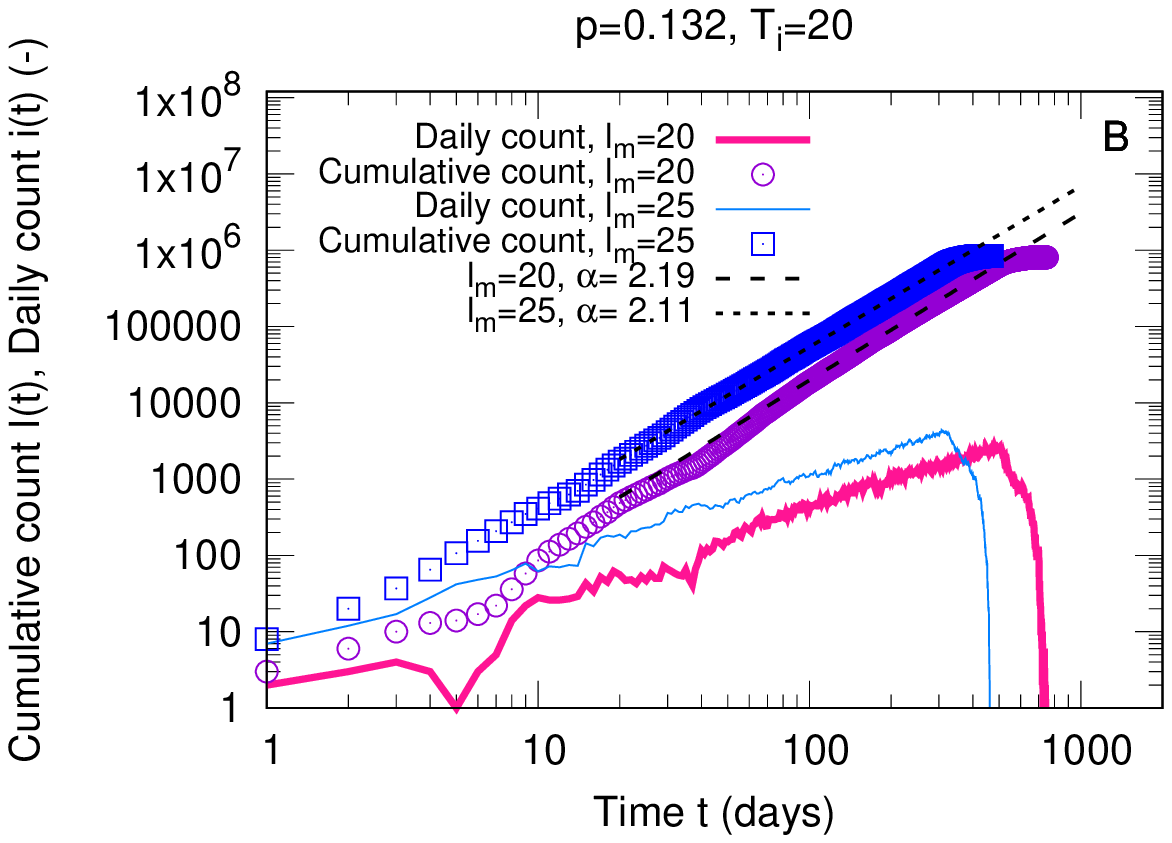}

\caption{\label{fig:simulation_2}The computer simulations of the effect of
the maximal path length $l_{m}$ and duration of infected state $T_{i}$
on the  exponents $\alpha$, on the total count of infected cases
$M$ and on the outbreak wave duration $T$. }
\end{figure}

If the probability of virus transmission $p=0.132,$ the duration
of infection state $T_{i}=14$ are constant and the maximal path length
$l_{m}$ decreases from $l_{m}=25$ to $l_{m}=15$, then the computer
simulations (Figure \ref{fig:simulation_2} (A) show a decrease of
the maximal count of infected cases $M$ and a decrease of the outbreak
wave duration $T$. The decrease of the maximal path length $l_{m}$
from $l_{m}=25$ to $l_{m}=20$, for parameters $p=0.132$ and $T_{i}=20$,
has opposite the effect on the maximal count of infected cases $M$
and the outbreak duration $T$. In this case, the outbreak duration
$T$ decreases and the maximal count of infected cases $M$ increases,
but $M<N$ ( (Figure \ref{fig:simulation_2} (B) ). 

The computer simulations, for the parameters $p=0.132$, $T_{i}=5$
and $90\leq l_{m}\leq120$, demonstrate the effects of the maximal
path length $l_{m}$ on the outbreak evolution (Supplementary Materials
(Figure S1)). The graphs show a decrease of the outbreak duration
$T$ that is correlated to the increase of the total count of infected
cases $M$, if the maximal path length $l_{m}$ increases. 

If we ran several computer simulations, for the constant parameters
$p=0.132$, $T_{i}=5$ and $l=90$ ((Supplementary Materials (Figure
S2))) we observed a complex growth dynamic. The Hurst exponents of
these time series are $H_{m}$, $0.42\leq H_{90}\leq0.61$. We observed
a random growth of the outbreaks and a change from anti-persistent
to persistent fluctuations $i(t)$, probably as the effect of a change
of the initial conditions and the complexity of dynamics. Such behaviors
are characteristic for complex systems that exhibit chaos. 

\section*{Discussion}

Many species that are exposed to pathogens can alter their behaviors
in ways to maximize benefits and minimize cost. Studies of social
behaviors of nonhuman animals have the potential to provide important
insights into ecological and evolutionary processes relevant to human
health, including pathogen transmission dynamics and virulence evolution
\cite{Stockmaiereabc8881}. 

Our aim has been to implement the most important feature of animals
to adapt to the pathogen enemy \cite{Stockmaiereabc8881} However,
we had to consider the human specific environmental and social conditions
that can dramatically change the outbreak evolution (see the model
in the Supplementary Materials). The fact that the individual mobility
patterns collapse into a single spatial probability distribution,
indicating that, despite the diversity of their travel history, humans
follow simple reproducible patterns \cite{Gonzalez2008} led us to
implement in the model a quenched spatio-temporal mobility pattern.
It is possible to identify individuals that have a high number of
daily personal contacts. In the time of pandemic, social behaviors
of these individuals are considered for the superspreading from an
epidemiological point of view. They have a potential to infect a large
number of their daily contacts \cite{Lipsitch_2003}. Superspreading
skews a distribution of the reproductive numbers $R$ \cite{Lipsitch_2003, Lloyd-Smith2005}.
We implemented supersperading as a variable length of daily path $m$
that is assigned to each individual, i.e. the node $n_{i,j}$. The
daily paths, $m$, follow a simple power law distribution \cite{Liljeros2001,Bettencourt1438,Schlapfer,Tizzoni2015}
(Supplementary Materials).

A transmission of the virus from an individual to an individual we
modeled as a branching process \cite{Antia2003}, where individuals
can move every day along the predestined paths (Supplementary Materials),
that are considered for the quenched mobility patterns \cite{Gonzalez2008}.
Recovered individuals are naturally concentrated in the spatio-temporal
structures along these quenched mobility patterns. Thus the individuals
create spatio-temporal barriers for the next infection contagion.
We note that a transition of individuals from the infected to the
recovered status represents a memory effect. In this way the evolution
of the spatio-temporal distribution of recovered individuals slows
down a virus spread, that can lead to spontaneous stop of virus contagion
without a need to infect the whole population as it is necessary in
the classical models \cite{Pastor_Satorras, Newman_2018, Barabasi_2016}.
Individual mobility patterns of the paths of the length $m$, where
the lengths, $m$, follow a power law distribution, were introduced
in the model to mimic the basic social and environmental human behaviors.

The results of computer simulations show the spontaneous stopping
of outbreak waves. The duration of outbreak $T$ and maximal size
of the outbreak $M$ are random variables that span broad intervals.
These quantities are independent of the population size $N$, $M\ll N$.
Such behaviors indicate the breaking of the scaling relationships
near the outbreak threshold \cite{Ben_Naim}. We think that the percolation
phenomenon should be one of the reasons for this broken scaling relationship.
However, spontaneous stopping of the outbreak, i.e. $M\ll N$, cannot
be simply explained only by considering the percolation phenomenon
\cite{Stauffer, Sahimi_2021}, because the percolation processes are
not obviously linked to the broad distributions of observable quantities.
The broad range of observable values of $M$ and $T$ together with
the spontaneous stopping of outbreak indicate that the outbreak growth
process could rather resemble the avalanche dynamics a near the self-organized
criticality \cite{Bak, Pruessner_2012}, if a probability of an occurrence
of the outbreak of size $M$ will follow power law $M^{-\tau},$ where
$\tau$ is a scaling exponent. The proposed model, in contrast to
the deterministic sandpile-like models \cite{Bak, Pruessner_2012},
has no flux of any conserved quantity. 

Forest Fire Model (FFM) of Bak, Chen and Tang and Drossel and Schwabl
\cite{Pruessner_2012}, are stochastic models that are linked to percolation
phenomenon \cite{Pruessner_2012}. Natural forest fires were directly
associated with the FFM \cite{Malamud}. Our model has features of
natural forest fires \cite{Malamud, Pruessner_2012}, where the time
evolution of forest fires is determined by the previous forest fires.
This important observation has been implemented in the model as a
spatio-temporal memory effect. 

We unexpectedly found that the model produces the fluctuations of
the daily count of infected cases $i(t)$ that show the Hurst exponents
$0.29\leq H\leq0.95$. A signature of chaotic behaviors (Figure S2
in the Supplementary Materials) has been observed. These features
of the model could be useful to explain complexity either outbreak
evolution or other similar natural systems. 

The strong effect of the initial conditions on the outbreak evolution
enables us to understand the benefits of early mobility restrictions
\cite{Cernak_2021, Brand} to change the outbreak dynamic (Figure
S2), in comparison with the effects of the intermediate measures that
are applied late, that are less intensive, that take longer time \cite{Kissler860}.
For example, in the Slovak Republic and in the Czech Republic (the
member states of European Union), the first SARS-CoV-2 wave has never
begun as a consequence of early and strong changes of the initial
conditions of the outbreak growth for a short time. However, in the
next SARS-CoV-2 outbreak waves the measures were weak and long \cite{Kissler860}
that led to the full development of the outbreak waves \cite{Cernak_2021},
a long term dissatisfaction in society and a dramatic increase of
the deaths.\textbf{ }After the first SARS-CoV-2 wave, the Slovak Republic
reported 54 total deaths on 01. October 2020. However, one year later
on 01. October 2021, before the third wave, 12660 total deaths were
reported.

In summary, we have investigated a computational model of virus contagion,
in which a spatio-temporal memory effect was introduced. The model
generates the time series of positive numbers that show the fluctuation
from the anti-persistent to persistent behaviors. A summation of these
numbers follows the power law with nontrivial exponents. We unexpectedly
observed a signature of chaotic behavior. A phase diagram of the model
is not known, thus it should be investigated in the future. The model
is possible to use in the studies of other natural phenomena that
show the time series of quantities with nontrivial Hurst exponents
i.e. $H\neq0.5$. The model could be easily updated to study actual
challenging tasks like reinfection, waning immunity \cite{Shaman527},
dominance of virus variants in population \cite{Callaway} or other
natural phenomena.

\nocite{*}
\printbibliography

\end{document}


\title{Supplementary Materials for \\ Spatio-temporal dynamics of outbreak
on a lattice with quenched mobility patterns}

\author{Jozef \v{C}ern\'{a}k\footnote{jozefcernak@gmail.com}}
\maketitle

\lyxaddress{Department of Nuclear and Sub-Nuclear Physics, Faculty of Science,
Institute of Physics, Pavol Jozef Šafárik University in Košice, Košice,
Slovak Republic}

\tableofcontents{}\newpage{}

\section{Model\label{sec:Model}}

\begin{figure}
\centering

\includegraphics[width=6cm]{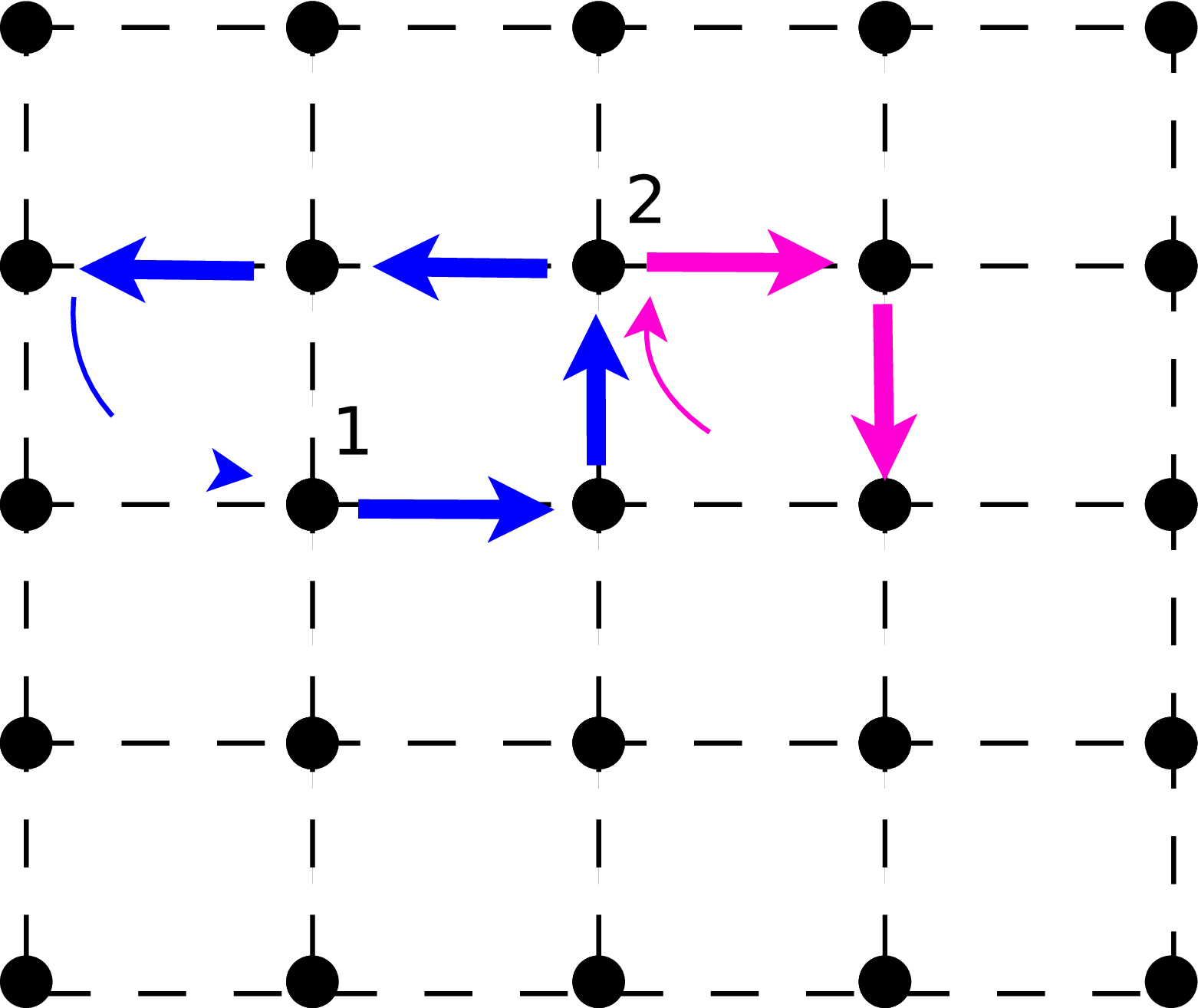}

\caption{\label{fig:Paths} The random paths of the length $m$ are generated
using the Markovian process (Eq. \ref{eq:1}). Two paths of the lengths,
$m=4$ and $m=2$, are shown. The paths constrain a movement of the
individuals, and they create a quenched disorder in the 2D lattice
$L\times L$ of the size $L$, where $L=5$. }
\end{figure}

The diffusion of individuals is defined on a two-dimensional (2D)
lattice of size $L\times L$, $L$ is a size in one dimension. An
individual is assigned to the lattice node $n_{ij}$.  The individuals
create a metapopulation of the size $N=L\times L$. The individual
can be mobile or immobile. The mobile individual may escape the node
$n_{ij}$ and diffuse, in contrast to the immobile individual that
stands  in the  node $n_{ij}$. We define the time $t$. If all mobile
individuals diffuse $m$ steps from the site $n_{ij}$ (Eq. \ref{eq:1})
then the time $t$ is increased by one unit $t=t+1$. After $m$ steps
each mobile individual returns to the initial node $n_{ij}$ (Figure
\ref{fig:Paths}). Mobile individuals are healthy or infected. Infected
individuals are infected for the time $T_{i}$, after a time $t>T_{i}$,
they become healthy but immobile. It is allowed only for mobile individuals
to interact and transmit the infection. A movement of an individual
is constrained in a stapce. They can diffuse only along the spatially
fixed random paths of the length $m$, $m\leq l_{m}$. The parameter
$l_{m}$ is a maximal length of the path. For each individual $n_{ij}$,
we define the path of the length, $m$, only once, at the simulation
initialization. The path is a set of $m$ nodes $n_{ij}$, where indices
$i$ and $j$ are determined using the Markovian process

\begin{eqnarray}
i_{k+1}={\displaystyle \sum_{k=0}^{m-1}i_{k}+a_{1},} & j_{k+1}={\displaystyle \sum_{k=1}^{m-1}j_{k}+a_{2}}, & a_{1,2}=\begin{cases}
1 & 0\leq\xi_{1,2}<\frac{1}{3}\\
0 & \frac{1}{3}\leq\xi_{1,2}<\frac{2}{3}\\
-1 & \frac{2}{3}\leq\xi_{1,2}<1
\end{cases},\label{eq:1}
\end{eqnarray}

$i_{k=0}=i$, $j_{k=0}=j$, $a_{1}$and $a_{2}$ are random integer
numbers and $\xi_{1}$ and $\xi_{2}$ are uncorrelated random variables
$0\leq\xi_{1,2}<1$ that are evaluated for each $k$. In simulations,
the periodic boundary conditions are used to conserve a number of
individuals $N$ on the lattice. We proposed that the distribution
of the path lengths $m$ follows the power law $P_{m}=m^{-1}$, where
$m\geq1$ and a path length $m\leq l_{m}$ (see the Discussion in
the main text).

Two individuals, $1$ and $2$, have forever defined the random paths
that are shown in Figure \ref{fig:Paths}. The arrows are returns
of the individuals in the initial nodes $n_{ij}$ after $m\text{=4}$
and $m=2$ steps.

A mobile individual interacts with other individuals, $n_{ij}$, that
lie on his daily path. An infection is transmitted from infected individuals
to susceptible individuals with a probability $p$ as a branching
process. The probability of infection transmission $p$, the infection
time $T_{i}$ and the maximal length $l_{m}$ are model parameters.

\section{Computational methods}

The computational model (Section \ref{sec:Model}) is implemented
in the Python language (``m5.py'' \cite{local_data}). We tested
the program ``m5.py'' using the OS Ubuntu 18.04.6 LTS. Type in the
terminal: \textsl{python3 m5.py} to run a computer simulation. The
program prints the time series: the time $t$, the daily count of
infected cases $i(t)$ and the cumulative count of infected cases
$I(t)$. 

For plotting and fitting the experimental and computer simulation
data we used a portable command-line driven graphing software the
Gnuplot \cite{gnuplot}. SageMath (a free open-source mathematics
software system licensed under the GPL) \cite{sagemath} has built
the algorithm to calculate the Hurst exponent \cite{Peters} as the
function \textit{hurst\_exponent().} 

\section{Data}

\subsection{Experimental data}

World Health Organization (WHO) collects the statistics of the COVID-19
pandemic, which are publicly available in the WHO archive \cite{who}.
Statistical data were downloaded from the link \cite{who_link}, as
the comma-separated values (CSV file format). Data columns: ``Date\_reported'',
``Country\_code'', ``New\_cases'', and ``Cumulative\_cases''
were exported from the Microsoft Excel data sheet ``WHO-COVID-19-global-data.xlsx''.
The file ``WHO-COVID-19-global-data.xlsx'' and exported data used
to plot the graphs in Figure 1 (the main text) are enclosed in the
Local Data Directories \cite{local_data}.

\subsection{Computer simulation data}

The results of computer simulations are stored in data files ``File.txt''
\cite{local_data}. The probability of infection transmission $p$,
the duration of infection window $T_{i}$, and the maximal path length
$l_{m}$ are the parameters that were used to generate the time series
of $i(t)$ and $I(t)$, that are stored in data files ``File.txt''.
Data files and the parameters $p$, $T_{i}$ and $l_{m}$ are arranged
in Tables \ref{tab:com_sim} and \ref{tab:graphs_lm}. These data
were used to plot the graphs in Figures 2, 3 (the main text), \ref{fig:Simulations-of-cumulative}
and \ref{fig:Outberak-wave-duration}.

Data files and the parameters $p$, $T_{i}$ and $l_{m}$ are arranged
in Tables \ref{tab:com_sim}. 

\begin{table}[H]
\begin{tabular}{|c|c|c|c|c|}
\hline 
$p$ & $T_{i}$ & $l_{m}$ & File (.txt) & Figure 2\tabularnewline
\hline 
\hline 
\multirow{5}{*}{$0.130$} & \multirow{17}{*}{$14$} & \multirow{17}{*}{$20$} & \textbf{v4} & \multirow{2}{*}{ (A)}\tabularnewline
\cline{4-4} 
 &  &  & \textbf{v40\_3} & \tabularnewline
\cline{4-5} 
 &  &  & v4\_1 & \multirow{3}{*}{-}\tabularnewline
\cline{4-4} 
 &  &  & v40\_1 & \tabularnewline
\cline{4-4} 
 &  &  & v40\_2 & \tabularnewline
\cline{1-1} \cline{4-5} 
\multirow{4}{*}{$0.132$} &  &  & \textbf{v5} & \multirow{2}{*}{(B)}\tabularnewline
\cline{4-4} 
 &  &  & \textbf{v50\_1} & \tabularnewline
\cline{4-5} 
 &  &  & v5\_1 & \multirow{2}{*}{-}\tabularnewline
\cline{4-4} 
 &  &  & v50\_2 & \tabularnewline
\cline{1-1} \cline{4-5} 
\multirow{4}{*}{$0.135$} &  &  & \textbf{v2} & \multirow{2}{*}{(C)}\tabularnewline
\cline{4-4} 
 &  &  & \textbf{v20\_2} & \tabularnewline
\cline{4-5} 
 &  &  & v2\_1 & \multirow{2}{*}{-}\tabularnewline
\cline{4-4} 
 &  &  & v20\_1 & \tabularnewline
\cline{1-1} \cline{4-5} 
\multirow{4}{*}{$0.140$} &  &  & \textbf{v1\_1} & \multirow{2}{*}{(D)}\tabularnewline
\cline{4-4} 
 &  &  & \textbf{v10\_2} & \tabularnewline
\cline{4-5} 
 &  &  & v1 & \multirow{2}{*}{-}\tabularnewline
\cline{4-4} 
 &  &  & v10\_1 & \tabularnewline
\hline 
\end{tabular} %
\begin{tabular}{|c|c|c|c|c|}
\hline 
$p$ & $T_{i}$ & $l_{m}$ & File (.txt) & Figure 3\tabularnewline
\hline 
\hline 
\multirow{17}{*}{$0.132$} & \multirow{9}{*}{$14$} & \multirow{6}{*}{$15$} & \textbf{v90\_5} & (A)\tabularnewline
\cline{4-5} 
 &  &  & v9\_5 & \multirow{5}{*}{-}\tabularnewline
\cline{4-4} 
 &  &  & v9\_3 & \tabularnewline
\cline{4-4} 
 &  &  & v9\_4 & \tabularnewline
\cline{4-4} 
 &  &  & v90\_1 & \tabularnewline
\cline{4-4} 
 &  &  & v90\_3 & \tabularnewline
\cline{3-5} 
 &  & \multirow{3}{*}{$25$} & \textbf{v80\_1} & (A)\tabularnewline
\cline{4-5} 
 &  &  & v8 & \multirow{2}{*}{-}\tabularnewline
\cline{4-4} 
 &  &  & v80\_2 & \tabularnewline
\cline{2-5} 
 & \multirow{8}{*}{$20$} & \multirow{4}{*}{$20$} & \textbf{v6\_1} & (B)\tabularnewline
\cline{4-5} 
 &  &  & v6 & \multirow{3}{*}{}\tabularnewline
\cline{4-4} 
 &  &  & v60\_1 & \tabularnewline
\cline{4-4} 
 &  &  & v60\_2 & \tabularnewline
\cline{3-5} 
 &  & \multirow{4}{*}{$25$} & \textbf{v7\_1} & (B)\tabularnewline
\cline{4-5} 
 &  &  & v7 & \multirow{3}{*}{}\tabularnewline
\cline{4-4} 
 &  &  & v70\_1 & \tabularnewline
\cline{4-4} 
 &  &  & v70\_2 & \tabularnewline
\hline 
\end{tabular}\caption{\label{tab:com_sim}The parameters $p$, $T_{i}$ and $l_{m}$, and
data files (``File.txt'') of the graphs in Figures 2 and 3 (the
main text).}
\end{table}

\begin{table}[H]
\begin{center}

\begin{tabular}{|c|c|c|c|c|}
\hline 
$p$ & $T_{i}$ & $l_{m}$ & File (.txt) & Figure\tabularnewline
\hline 
\hline 
\multirow{16}{*}{$0.132$} & \multirow{16}{*}{$5$} & \multirow{4}{*}{$90$} & \textbf{5\_90\_0132} & \ref{fig:Simulations-of-cumulative}, \ref{fig:Outberak-wave-duration}\tabularnewline
\cline{4-5} 
 &  &  & 5\_90\_0132\_1 & \ref{fig:Outberak-wave-duration}\tabularnewline
\cline{4-5} 
 &  &  & 5\_90\_0132\_2 & \ref{fig:Outberak-wave-duration}\tabularnewline
\cline{4-5} 
 &  &  & 5\_90\_0132\_3 & \ref{fig:Outberak-wave-duration}\tabularnewline
\cline{3-5} 
 &  & \multirow{4}{*}{$100$} & \textbf{5\_100\_0132} & \ref{fig:Simulations-of-cumulative}\tabularnewline
\cline{4-5} 
 &  &  & 5\_100\_0132\_1 & \multirow{3}{*}{-}\tabularnewline
\cline{4-4} 
 &  &  & 5\_100\_0132\_2 & \tabularnewline
\cline{4-4} 
 &  &  & 5\_100\_0132\_3 & \tabularnewline
\cline{3-5} 
 &  & \multirow{4}{*}{$110$} & \textbf{5\_110\_0132} & \ref{fig:Simulations-of-cumulative}\tabularnewline
\cline{4-5} 
 &  &  & 5\_110\_0132\_1 & \multirow{3}{*}{}\tabularnewline
\cline{4-4} 
 &  &  & 5\_110\_0132\_2 & \tabularnewline
\cline{4-4} 
 &  &  & 5\_110\_0132\_3 & \tabularnewline
\cline{3-5} 
 &  & \multirow{4}{*}{$120$} & \textbf{5\_120\_0132} & \ref{fig:Simulations-of-cumulative}\tabularnewline
\cline{4-5} 
 &  &  & 5\_120\_0132\_1 & \multirow{3}{*}{}\tabularnewline
\cline{4-4} 
 &  &  & 5\_120\_0132\_2 & \tabularnewline
\cline{4-4} 
 &  &  & 5\_120\_0132\_3 & \tabularnewline
\hline 
\end{tabular} 

\end{center}

\caption{\label{tab:graphs_lm} The probability of infection transmission $p=0.132$,
the duration of infection window $T_{i}=5$, and the maximal path
lengths $l_{m}=90$, $100$, $110$ and $120$ are parameters. Data
files (``File.txt'') were used to plot the graphs in Figures \ref{fig:Simulations-of-cumulative}
and \ref{fig:Outberak-wave-duration} and to determine the Hurst exponents.}
\end{table}

\section{Results}

\begin{figure}[H]
\includegraphics[width=8cm]{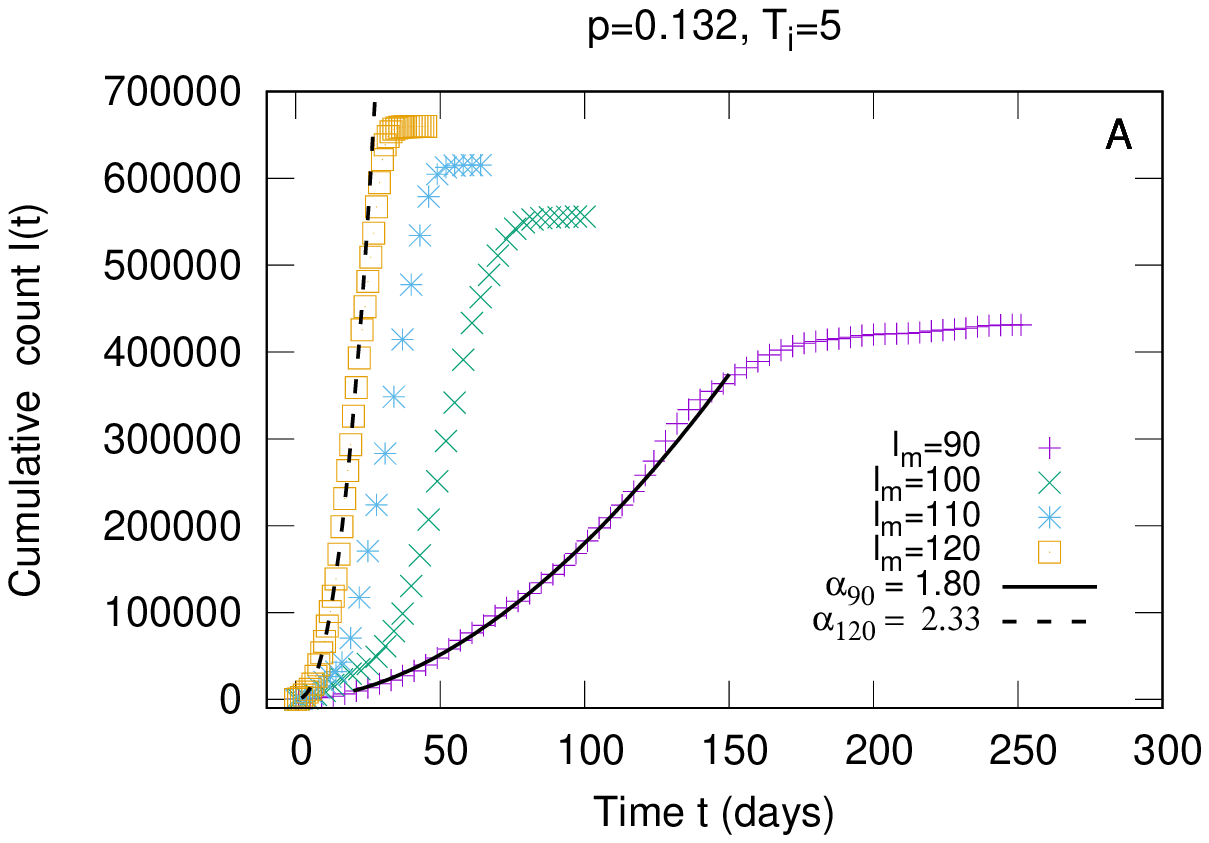}\includegraphics[width=8cm]{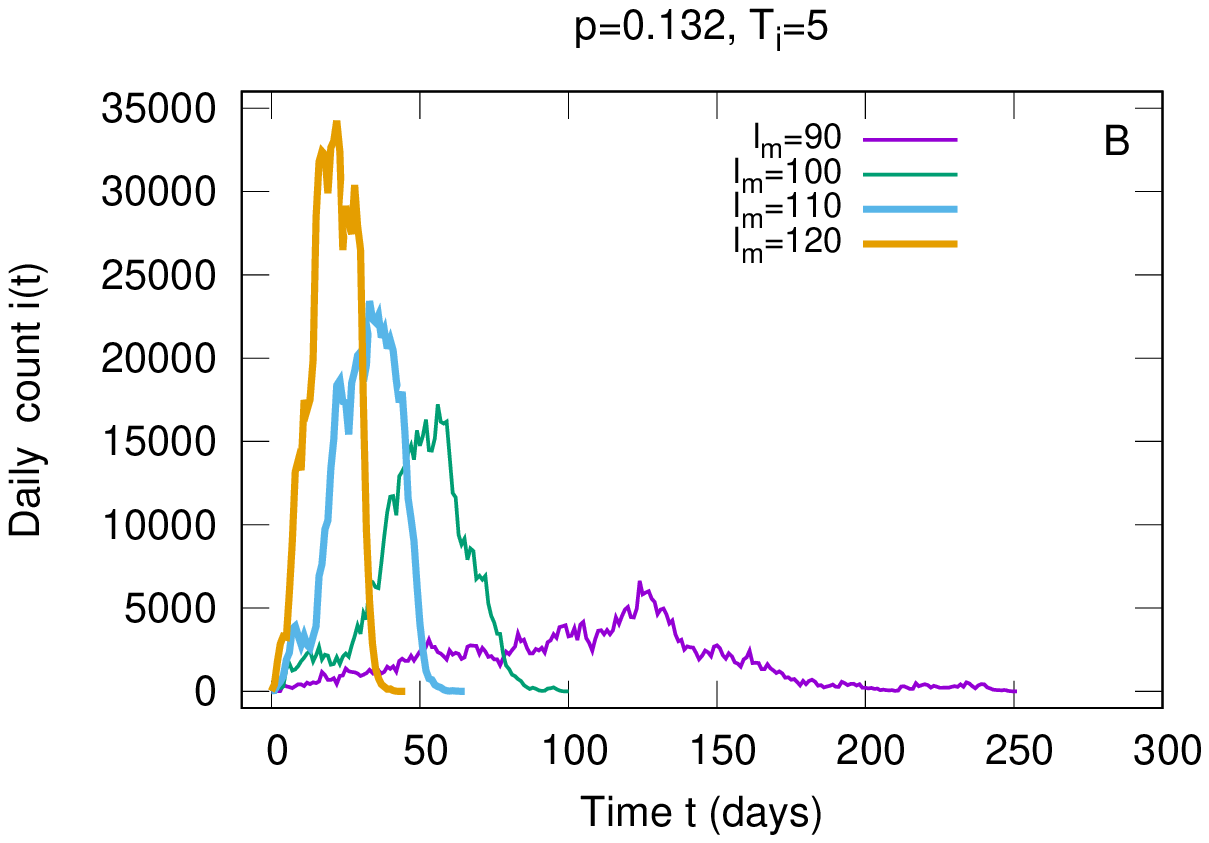}

\caption{\label{fig:Simulations-of-cumulative}The computer simulations of
the cumulative count of infected cases $I(t)$ (A) and daily count
of infected cases $i(t)$ (B), for fixed parameters: $p=0.132$, $T_{i}=5$,
and the variable parameter $l_{m}$, $l_{m}=90$, $100$, $110$ and
$120$.}
\end{figure}

Simulations are initiated using only one infected case, that it is
placed randomly on the lattice. In such case, a probability to start
an outbreak may be low \cite{Lipsitch_2003}. We had to run the initialization
of simulation many times for certain parameters. We observed a spontaneous
stopping of the outbreak after the time $T$, for all parameters $p$,
$T_{i}$and $l_{m}$ (Figure \ref{fig:Simulations-of-cumulative}),
i.e. the total count of infected cases $M$ was always smaller than
a population size $N$, i.e. $M<N$, and $N=1\times10^{6}$. For constant
parameters $p=0.132$ and $T_{i}=5$, computer simulations (Figure
\ref{fig:Simulations-of-cumulative}) show the effect of an increase
of $l_{m}$ on the increase of the total count of infected cases $M$
(Figure \ref{fig:Simulations-of-cumulative} (A)) and on the decrease
of the outbreak wave duration $T$ ((Figure \ref{fig:Simulations-of-cumulative})
(B)).

\begin{figure}[h]
\includegraphics[width=8cm]{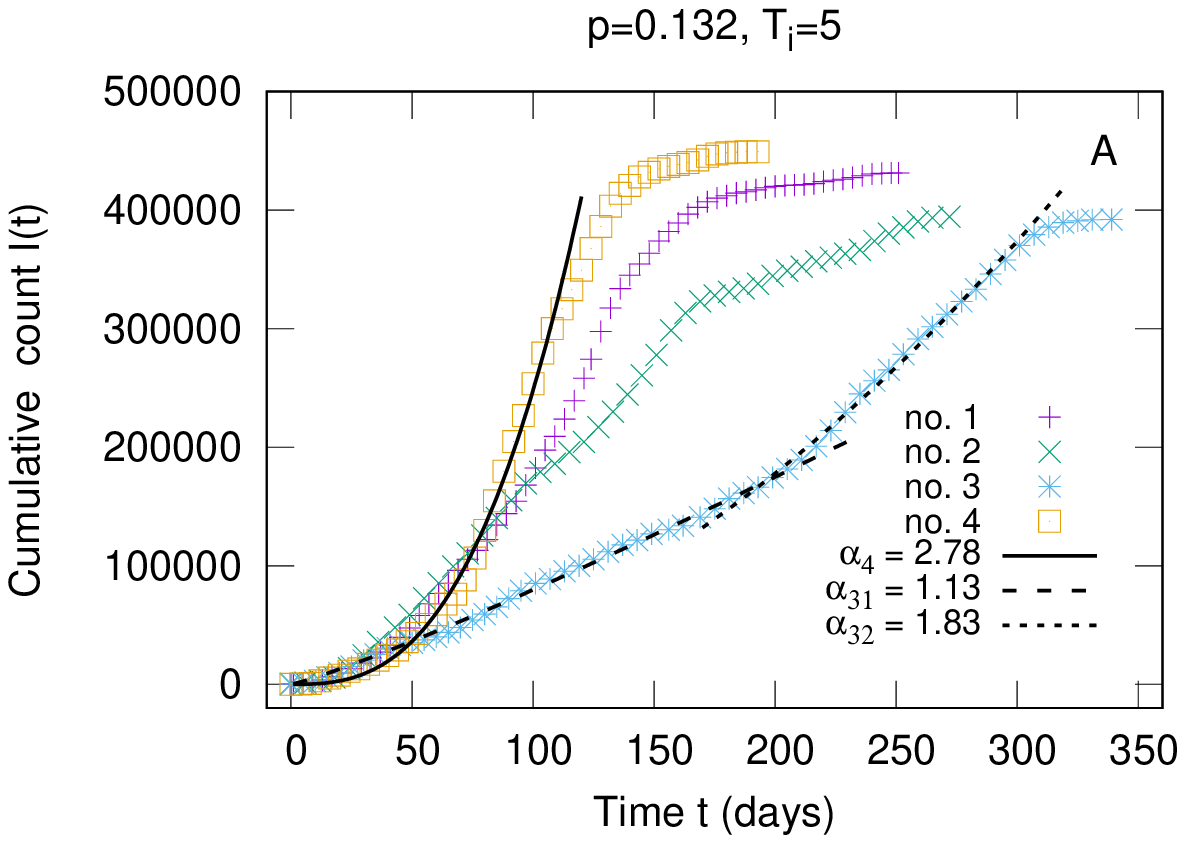}\includegraphics[width=8cm]{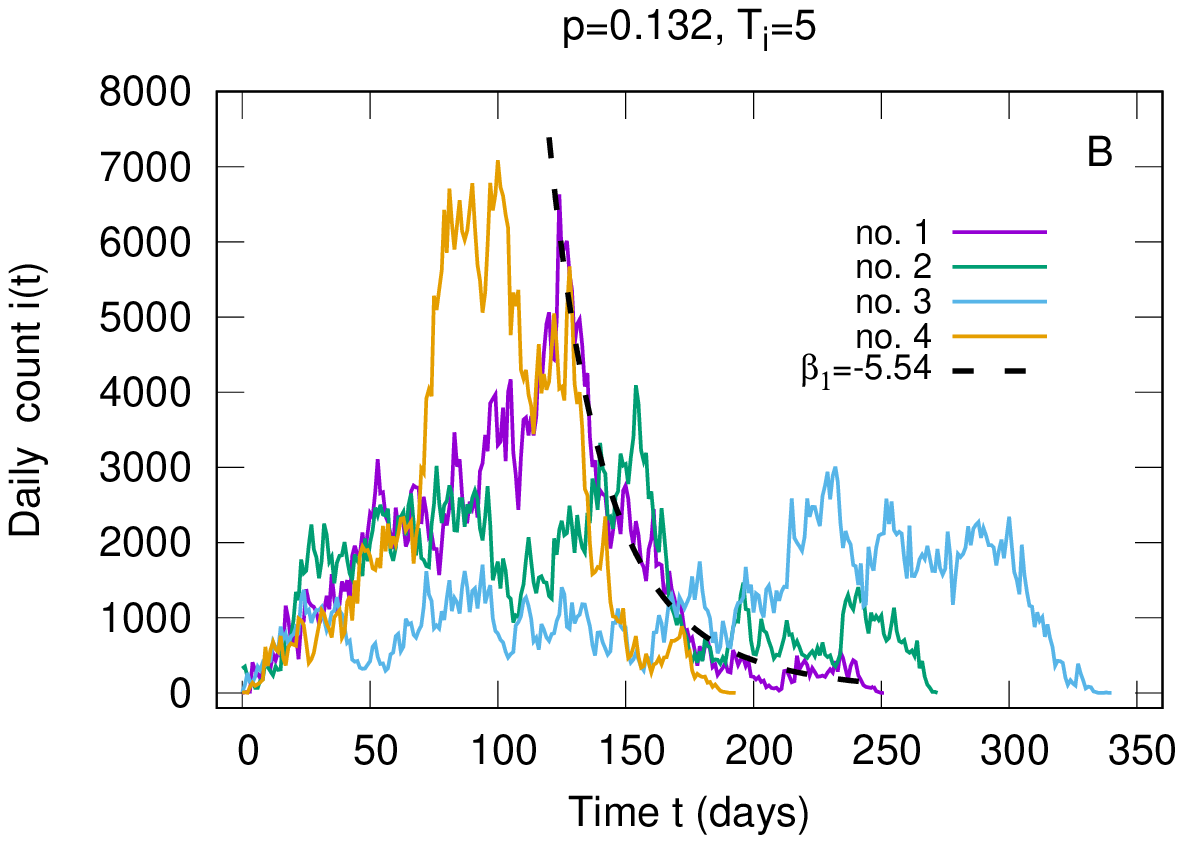}\caption{\label{fig:Outberak-wave-duration}The computer simulations for the
fixed parameters: $p=0.132$, $T_{i}=5$, and $l_{m}=90$ (see Table
\ref{tab:com_sim}). The exponents, $\alpha$ and $\beta$, are of
the power laws $i(t)\sim t^{\beta}$ and $I(t)\sim t^{\alpha}$.}
\end{figure}

We observed a stochastic growth process of the cumulative number of
infected cases $I(t)$ (Figure \ref{fig:Outberak-wave-duration}),
for the constant parameters $p=0.132$, $T_{i}=5$ and $l_{m}=90$,
that led to the spontaneous stopping of a virus contagion ($M<N$)
and to the random durations of the outbreak waves, $T$.

We observed that, the growth of the outbreak and its decay follow
the power laws $i(t)\sim t^{\beta}$ and $I(t)\sim^{\alpha}$(Figures
\ref{fig:Simulations-of-cumulative} (A) and \ref{fig:Outberak-wave-duration}),
where the exponents are $1.13\leq\alpha\leq2.28$ and $\beta=-5.54$.

For computer simuations (Figure 2 in the main text and Table \ref{tab:com_sim}),
with the constant parameters $T_{i}=14$ and $l_{m}=20$, and the
variable probability of infection transmission $p$, $0.130\leq p\leq0.140$,
we determined the Hurst exponents $H$ using the fluctuations of the
daily count of infected cases $i(t)$ as: Figure 2 (A) $0.39\leq H\leq0.46$,
Figure 2 (B) $0.29\leq H\leq0.44,$ Figure 2 (C) $0.43\leq H\leq0.62$,
and Figure 2 (D) $0.42\leq H\leq0.45$.

For computer simulation, with the constant probability of infection
transmission $p=0.132$ and the variable parameters $T_{i}=14$ and
$20$, and $l_{m}=15$, $20$ and $25$ (Figure 3 in the main text
and Table \ref{tab:com_sim}), the Hurst exponents $H$ of the daily
count of infected cases $i(t)$ were determined as: Figure 3 (A) $0.29\leq H$$\leq0.43$
($T_{i}=14$ and $l_{m}=15$), and $0.41\leq H$$\leq0.49$ ($T=14$
and $l_{m}=25$), Figure 3 (B) $0.43\leq H\leq0.54$ ($T_{i}=20$
and $l_{m}=20$), and $0.36\leq H$$\leq0.48$ ($T_{i}=20$ and $l_{m}=25$).

The fluctuations of daily count infected cases $i(t)$, for parameters
and data files that arranged in Table \ref{tab:graphs_lm}, have the
Hurst exponents $H_{m}$: $0.42\leq H_{90}\leq0.61$ (a maximal duration
of waves, $193\leq T\leq340$), $0.30\leq H_{100}\leq0.77$ ($98\leq T$$\leq112$),
$0.44\leq H_{110}\leq0.95$ ($60\leq T\leq$71) and $0.77\leq H_{120}\leq0.97$
($43\leq T\leq45$).

The Hurst exponents $H_{m}$ of the fluctuation of daily count of
infected cases $i(t)$ in Figure \ref{fig:Simulations-of-cumulative}
(Table \ref{tab:graphs_lm}) are: $H_{90}=0.55$, $H_{100}=0.77$,
$H_{110}=0.90$ and $H_{120}=0.77$.

\nocite{*}
\printbibliography